\documentclass[prd,twocolumn,aps,nofootinbib,eqsecnum,notitlepage,nobibnotes,superscriptaddress,preprintnumbers,floatfix]{revtex4-1}
\usepackage{subfig}
\usepackage[a4paper, left=20mm,right=20mm,top=20mm,bottom=20mm]{geometry}

\usepackage{mathtools}
\usepackage{tensor}
\usepackage{amsmath, amsfonts,bm}
\usepackage{derivative}
\usepackage{bm}
\usepackage{amssymb}

\usepackage{suffix}

\usepackage{amsmath, amsfonts,bm}
\usepackage{color}
\usepackage{graphicx}
\usepackage{subfig}
\definecolor{myred}{rgb}{0.9, 0.17, 0.31}

\usepackage[a4paper, left=20mm,right=20mm,top=20mm,bottom=20mm]{geometry}
\usepackage[colorlinks=true,hyperfootnotes=true, linkcolor=myred, citecolor=cyan, urlcolor=magenta]{hyperref}

\newcommand\mathcomma{\,,}
\newcommand\mathperiod{\,.}

\DeclareMathAlphabet{\mathup}{OT1}{\familydefault}{m}{n}

\def\dd{\mathrm{d}}

\newcommand{\be}{\begin{equation}} 
\newcommand{\ee}{\end{equation}}

\begin{document}

\title{Is there evidence for CIDER in the Universe?}

\author{Bruno J. Barros}
\email{cstbru002@myuct.ac.za}
\affiliation{Cosmology and Gravity Group, Department of Mathematics and Applied Mathematics, University of Cape Town, Rondebosch 7700, Cape Town, South Africa}
\author{Diogo Castel\~ao}
\affiliation{Instituto de Astrof\'isica e Ci\^encias do Espa\c{c}o,\\ 
Faculdade de Ci\^encias da Universidade de Lisboa,  \\ Campo Grande, PT1749-016 
Lisboa, Portugal}
\author{Vitor da Fonseca}
\affiliation{Instituto de Astrof\'isica e Ci\^encias do Espa\c{c}o,\\ 
Faculdade de Ci\^encias da Universidade de Lisboa,  \\ Campo Grande, PT1749-016 
Lisboa, Portugal}
\author{Tiago Barreiro}
\affiliation{Instituto de Astrof\'isica e Ci\^encias do Espa\c{c}o,\\ 
Faculdade de Ci\^encias da Universidade de Lisboa,  \\ Campo Grande, PT1749-016 
Lisboa, Portugal}
\affiliation{ECEO, Universidade Lus\'ofona de Humanidades e 
Tecnologias, Campo Grande, 376,  PT1749-024 Lisboa, Portugal}
\author{Nelson J. Nunes}
\affiliation{Instituto de Astrof\'isica e Ci\^encias do Espa\c{c}o,\\ 
Faculdade de Ci\^encias da Universidade de Lisboa,  \\ Campo Grande, PT1749-016 
Lisboa, Portugal}
\author{Ismael Tereno}
\affiliation{Instituto de Astrof\'isica e Ci\^encias do Espa\c{c}o,\\ 
Faculdade de Ci\^encias da Universidade de Lisboa,  \\ Campo Grande, PT1749-016 
Lisboa, Portugal}

\begin{abstract}
In this work we analyze the full linear behaviour of the constrained interacting dark energy (CIDER) model, which is a conformally coupled quintessence model tailored to mimic a $\Lambda$CDM expansion. We compute the matter and temperature anisotropies power spectra and test the model against recent observational data. We shed light on some particular subtleties of the background behaviour that were not fully captured in previous works, and study the physics of the linear cosmological observables. One novelty found was that matter perturbations are enhanced at large scales when compared with the ones of the standard $\Lambda$CDM. The reason and impact of this trend on the cosmological observables and on the physics of the early Universe are considered. We find that the introduction of the coupling parameter alleviates the $\sigma_8$ tension between early and late time probes although Planck data favours the $\Lambda$CDM limit of the model.
\end{abstract}

\maketitle
%%%%%%%%%%%%%%%%%%%%%%%%%%%%%%%%%%%%%%%%%%%%%%%%%%%%%%%%%%%%%%%%%%%%%%
%\twocolumngrid
%%%%%%%%%%%%%%%%%%%%%%%%%%%%%%%%%%%%%%%%%%%%%%%%%%%%%%%%%%%%%%%%%%%%%%
\section{Introduction}\label{sec:intro}
%%%%%%%%%%%%%%%%%%%%%%%%%%%%%%%%%%%%%%%%%%%%%%%%%%%%%%%%%%%%%%%%%%%%%%

 Over the last years cosmologists have been challenged with the existence of discrepancies among crucial cosmological parameters of the $\Lambda$CDM cosmology, which otherwise provides a remarkable fit to observational data. There are unexplained tensions between early and late time experiments in the current rate of expansion $H_0$ \cite{Mortsell:2018mfj,DiValentino:2019jae,DiValentino:2020zio,Vagnozzi:2019ezj,Efstathiou:2013via} and in the amplitude of the linear matter power spectrum $\sigma_8$ \cite{Macaulay:2013swa,DiValentino:2020vvd,Benisty:2020kdt,Douspis:2018xlj,Battye:2014qga}. In order to circumvent the former problem, theorists contemplate extensions beyond the standard model that assume a different dynamical evolution for the Hubble parameter $H(z)$ to try to better fit the data and alleviate this tension \cite{DiValentino:2021izs,Murgia:2020ryi,Pandey:2019plg}. On the other hand, it is also possible to consider models that change the linear behaviour of matter fluctuations to achieve the same end regarding the $\sigma_8$ observational puzzle \cite{DiValentino:2017oaw,FrancoAbellan:2020xnr,Barros:2018efl}. Within the plethora of extended models, one enticing possibility is provided by quintessence models \cite{Zlatev:1998tr,Copeland:2006wr,Copeland:1997et}, first proposed as the {\it cosmon} field \cite{Peccei:1987mm,Wetterich:1994bg}. Since in these theories the role of dark energy is played by a dynamical scalar field, instead of Einstein's cosmological constant $\Lambda$, the evolution of the scale factor of the Universe differs from the standard $\Lambda$CDM model and the $H_0$ tension can be addressed. Additionally, if one assumes couplings of this scalar source to the matter fields \cite{Barros:2019rdv,Amendola:2003wa,Amendola:1999qq,Kase:2019veo,vandeBruck:2020fjo,Dusoye:2020wom,Leithes:2016xyh,DAFONSECA2022100940}, the evolution for the matter density contrast, $\delta_m$, and thus the predictions for the observable $\sigma_8$, will inevitably differ from the $\Lambda$CDM ones. Therefore, one is able to tackle the observed cosmological $\sigma_8$ tension with scalar field models of dark energy, in particular with coupled quintessence. A linear conformal coupling between the quintessence field and matter was introduced in \cite{Amendola:1999er} for a specific exponential potential. The author explored the background behaviour of the model and the influence of the interaction on the overall cosmology, such as on the first acoustic peak of the cosmic microwave background (CMB), from which an upper bound on the coupling was found. Modified gravity theories can also produce expansion rates and matter perturbations that deviate from the standard model ones and alleviate the tensions \cite{DAgostino:2020dhv,Odintsov:2020qzd,Belgacem:2017cqo,Awad:2017yod,Bengochea:2010sg}. In some of those models, however, the extensions from $\Lambda$CDM arise from higher-order curvature terms, which may ultimately be interpreted as an effective fluid, {\it i.e.} dark energy. The possibility of distinguishing modified gravity and pure dark energy models was discussed in, {\it e.g.}, Refs.~\cite{Joyce:2016vqv,Amendola:2019xqj}.

In \cite{Barros:2018efl}, the authors have proposed a specific model in the form of a coupled scalar field $\phi$ that seems to be promising in alleviating the $\sigma_8$ tension. The model is tailored to mimic a $\Lambda$CDM expansion rate at background level, thus not tackling the $H_0$ tension between the CMB and late time observations since its distinct observational signatures only arise at linear level. Note that in Ref.~\cite{Asghari:2019qld} the authors also construct an interacting quintessence-dark matter model with a fixed background. However this is done in a different way by imposing that the coupling depends on the relative motion (velocities) of the dark components, also leading to deviations only at the level of perturbations (see also \cite{Figueruelo:2021elm}). Other theories that feature a background evolution identical or very similar to the standard model can be found in Refs.~\cite{Simpson:2010vh,Pourtsidou:2013nha,Baldi:2016zom}.

The small scale late time behaviour of matter perturbations were thoroughly analysed and tested against redshift space distortions (RSD) data in \cite{Barros:2018efl}. The authors have found that the coupling between the scalar field and dark matter (DM) suppresses the DM fluctuations, which inevitably slows down the clustering rate of matter. In Ref.~\cite{Barros:2019hsk} the authors explored the nonlinear regime of the model by evolving the second order matter perturbations. Since the model has a slower clustering rate for stronger dark energy-dark matter interactions, the collapse of matter perturbation spherical regions will be delayed in the cosmic history. Thus a higher amount of density contrast is required for a spherical region to collapse and form a bound structure. The predictions for the number of bound structures were computed, and the sensitivity of current missions to identify a non-zero value of the coupling was estimated. More recently, in \cite{Baldi:2022uwb} the author performed N-body simulations to study the physics of the present model at non-linear scales. The coupling induces a low-k suppression of the matter clustering in accordance with the finding of \cite{Barros:2018efl,Barros:2019hsk}. Moreover, accounting for non-linear corrections, the model suppresses halo abundances and inner densities. Finally, it was shown that the coupling strongly impacts the abundance of cosmic voids due to the slower growth of dark matter fluctuations.

Here, we complement the work carried out in \cite{Barros:2018efl,Barros:2019hsk,Baldi:2022uwb}, by evolving the full set of linear equations in the CIDER model with the Einstein-Boltzmann code CLASS \cite{class} to compute the matter and temperature angular power spectra as observables. We are also able to shed light on some subtleties of the background behaviour that were not fully captured in previous studies, and analyse the large scale demeanor of matter perturbations relating to the physics of the early Universe. We test the model with current weak lensing observations, complementing the analysis of \cite{Barros:2018efl} at low redshift, as well as with CMB data to further the analysis at high redshift. The background description of the model and its linear behaviour can be found in Sec.~\ref{sec:model} and Sec.~\ref{sec:pert} respectively. The parameter inference is reported in Sec.~\ref{sec:obs} and we conclude on the $\sigma_8$ tension in Sec.~\ref{sec:conclusions}.

%%%%%%%%%%%%%%%%%%%%%%%%%%%%%%%%%%%%%%%%%%%%%%%%%%%%%%%%%%%%%%%%%%%%%%
\section{Model}\label{sec:model}
%%%%%%%%%%%%%%%%%%%%%%%%%%%%%%%%%%%%%%%%%%%%%%%%%%%%%%%%%%%%%%%%%%%%%%

On a Friedmann-Lema\^itre-Roberson-Walker (FLRW) Universe, the equations governing the background evolution of the coupled species -- the dark energy scalar field $\phi$ with mean energy density ${\rho_{\phi}=\dot{\phi}^2/2a^2+V_{\phi}}$ and dark matter with mean energy density $\rho_c$ -- for coupled quintessence are well established in the literature \cite{Amendola:1999er,Amendola:2001rc} and read
\begin{eqnarray}
\ddot{\phi}+2aH\dot{\phi}+a^2V_{\phi}&=&a^2\kappa\beta\rho_c\mathcomma\label{KG}  \\
\dot{\rho}_c+3aH\rho_c &=& -\kappa\beta\dot{\phi}\rho_c\mathcomma\label{continuity}
\end{eqnarray}
where a dot denotes a derivative with respect to conformal time, $\tau$, ${H=\dot{a}/a^2}$ is the Hubble parameter, function of the scale factor $a$, $V_{\phi}=d V/ d \phi$ is the derivative of the potential with respect to the quintessence field $\phi$, the parameter $\beta$ quantifies the strength of the interaction between dark energy and dark matter, and ${\kappa^2=8\pi G}$. The Friedmann equation has the standard form,
\be\label{friedmann}
\frac{3}{\kappa^2}H^2= \sum_i \rho_i \mathcomma
\ee
enclosing the energy density of the coupled species, and standard non-interacting baryons and radiation, $\rho_b$ and $\rho_r$ respectively. 

The main feature of the CIDER model, in contrast to standard coupled quintessence, is the constraint relation
\be\label{assumption}
H=H_{\Lambda{\rm CDM}}\mathcomma
\ee
where $H_{\Lambda{\rm CDM}}$ depends on the energy densities of the cosmological constant, $\rho_{\Lambda}$, standard CDM matter, $\rho_{cdm}$, baryons and radiation.
This constraint ensures the $\Lambda$CDM expansion is reproduced; the potential has consequently the following form:
\be\label{potential}
V = \frac{\dot{\phi}^2}{2a^2}+\rho_{\Lambda}\mathperiod
\ee

From Eqs.~\eqref{assumption} and \eqref{potential} we may write the energy densities of quintessence and coupled dark matter as:
\begin{eqnarray}
\rho_{\phi} &=& \rho_{\Lambda}+\frac{\dot{\phi}^2}{a^2}\mathcomma \label{rho_phi} \\
\rho_c &=& \rho_{cdm} - \frac{\dot{\phi}^2}{a^2}\mathcomma\label{rho_c}
\end{eqnarray}
where $\rho_{cdm}=\rho_{cdm}^0a^{-3}$. Direct integration of Eq.~\eqref{continuity} gives
\be\label{rho_c_integrated}
\rho_c=\rho_{cdm} e^{-\kappa\beta\phi}
=\rho_{cdm}^0a^{-3}e^{-\kappa\beta\phi}\mathcomma
\ee
by fixing $\phi=0$ when $\rho_c = \rho_{cdm}$ with no loss of generality. With this choice, throughout the evolution $\beta\phi\geqslant0$ since ${\rho_c\leqslant\rho_{cdm}}$. Moreover, from Eq.~\eqref{rho_c}, at this time $\dot{\phi}=0$. From then on the energy transfer happens from the dark matter component into the scalar field, {\it i.e.} ${\beta\dot{\phi}\geqslant0}$. Focusing on the $\beta>0$ case, the equations being symmetric in $(\dot{\phi},\beta)$, $\phi$ has to grow to remain positive and therefore $\dot{\phi}\geqslant 0$.

\begin{figure}
    \centering
    \includegraphics[scale=0.53]{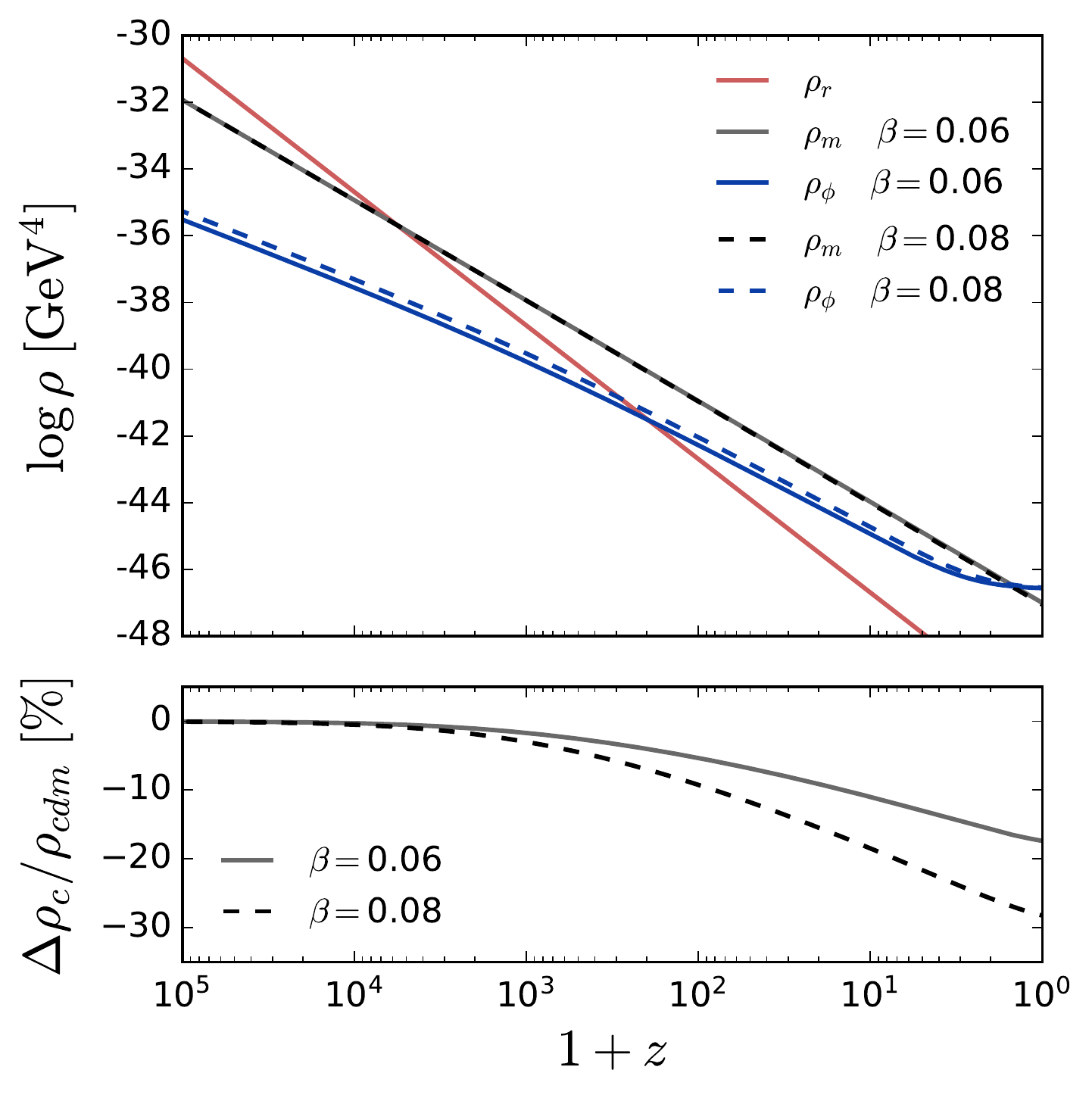}
    \caption{Evolution of the energy densities, $\rho_r$, ${\rho_m=\rho_c+\rho_b}$ and $\rho_{\phi}$, for two values of the coupling constant $\beta$ (top panel) and the relative difference of the evolution of the coupled dark matter density with respect to $\Lambda$CDM for two values of $\beta$ (bottom panel).}
    \label{fig:density_dilution}
\end{figure}

Using Eqs~\eqref{rho_c} and \eqref{rho_c_integrated}, one can write the following equation valid in any epoch,
\be\label{always_valid}
        e^{-\kappa\beta\phi}=1-\frac{\kappa^2\dot{\phi}^2}{3H_0^2\Omega_{cdm}^0}a\mathperiod
\ee
Let us now consider the radiation dominated era, where ${a=H_0\sqrt{\Omega_r^0}\tau}$ and Eq.\eqref{always_valid} becomes
\be
        e^{-\kappa\beta\phi}=1-\kappa^2\dot{\phi}^2\frac{\sqrt{\Omega_r^0}}{3H_0\Omega_{cdm}^0}\tau\mathcomma
\ee
and reduces to
\be
\label{reduced}
        \beta\kappa\phi-\kappa^2\dot{\phi}^2\frac{\sqrt{\Omega_r^0}}{3H_0\Omega_{cdm}^0}\tau=0\mathcomma
\ee
as long as $\phi$ is still sufficiently small to make the approximation {$e^{-\kappa\beta\phi}\simeq1-\kappa\beta\phi$}. Eq.~\eqref{reduced} possesses the following constant solution for $\dot{\phi}$:
\be
        \kappa\dot{\phi}=3\beta H_0\frac{\Omega_{cdm}^0}{\sqrt{\Omega_r^0}}\mathperiod
        \label{phi_rad}
\ee
From Eqs.~\eqref{rho_phi} and \eqref{phi_rad}, the energy density of the scalar field dilutes as $\rho_\phi\propto a^{-2}$ during the radiation dominated epoch. This is illustrated in Fig.~\ref{fig:density_dilution} computed with a modified version of the Einstein-Boltzmann code CLASS. Interestingly, in the numerical computation, the initial condition $\dot{\phi}_i$ can be set to the constant defined by  Eq.~\eqref{phi_rad}, and $\phi_i$ set to the corresponding value obtained from Eq.~\eqref{always_valid}, to immediately start at the radiation attractor solution.
%
%%%%%%%%%%%%%%%%%%%%%%%%%%%%%%%%%%%%%%%%%%%%%%%%%%%%%%%%%%%%%%%%%%%%%%
\section{Linear cosmological perturbations}\label{sec:pert}
%%%%%%%%%%%%%%%%%%%%%%%%%%%%%%%%%%%%%%%%%%%%%%%%%%%%%%%%%%%%%%%%%%%%%%
Let us now turn our attention to the linear behaviour of the present model. Accordingly, let us consider small perturbations along our FLRW background geometry in the Newtonian gauge, {\it i.e.},
\be\label{newt_gauge}
 \dd s^2 = a^2(\tau)\left[ -\left( 1+2\Psi \right)\dd \tau^2 + \left( 1-2\Phi \right)\delta_{ij}\dd x^i \dd x^j \right] \mathcomma
\ee
where $\Psi$ and $\Phi$ are the standard Bardeen potentials. The equation governing the evolution of interacting dark matter density contrast, $\delta_c$, is well known \cite{Barros:2018efl,Amendola:2003wa,Amendola:2014kwa} and reads,
\be\label{eq_delta_c}
\dot{\delta}_c+\theta_c-3\dot{\Phi} + \kappa\beta\dot{\delta\phi} =0\mathcomma
\ee
with $\theta_c$ being the perturbation on the dark matter velocity divergence, which evolve as,
\be
\dot{\theta}_c+\theta_c\left(aH-\kappa\beta\dot{\phi}\right)-k^2\Psi+\kappa\beta k^2\delta\phi=0\mathcomma
\ee
and the first order coupled Klein-Gordon equation giving the evolution for the scalar field perturbation, $\delta\phi$,
\begin{gather}
\ddot{\delta\phi} + 2aH\dot{\delta\phi} + \left( a^2V_{\phi\phi} +k^2 \right)\delta\phi-\left(\dot{\Psi}+3\dot{\Phi}\right)\dot{\phi} \nonumber \\
+2a^2\Psi V_{\phi} - a^2 \kappa\beta\rho_c\delta_c-2a^2\kappa\beta\rho_c\Psi=0\mathperiod
\end{gather}

Now, for our present model, using Eq.~\eqref{potential}, we find
\begin{eqnarray}
 V_{\phi} &=& -\frac{3}{2a}H\dot{\phi} +\frac{\kappa}{2}\beta \rho_c \\
V_{\phi\phi}&=&  -\frac{3}{2a}\dot{H}+\frac{9}{4}H^2 - \frac{\kappa\beta}{2}\rho_c\left( \kappa\beta+\frac{9}{2}\frac{aH}{\dot{\phi}} \right)
\label{Vphiphi}
\end{eqnarray}

We have modified the CLASS code by implementing the above coupled equations that govern the evolution of the perturbations in order to numerically predict the power spectrum of matter and the angular power spectrum of the cosmic microwave background (CMB).  We use them as observables in the next section to constrain the parameters of the CIDER model. In the simulations below, the $\Lambda$CDM parameters are fixed to Planck values \cite{Planck:2018vyg}, allowing to discriminate the response of the power spectra to the values of the interaction parameter $\beta$.

In \cite{Barros:2018efl}, the impact of the coupling between the dark species on the matter perturbations was studied in the Newtonian limit, thus capturing only the late time behaviour within the small scale regime. Solving the full linear equations resorting to a Boltzmann code allows us to shed light on the physics of the full scale linear regime.

\begin{figure}
    \centering
    \includegraphics[scale=0.4]{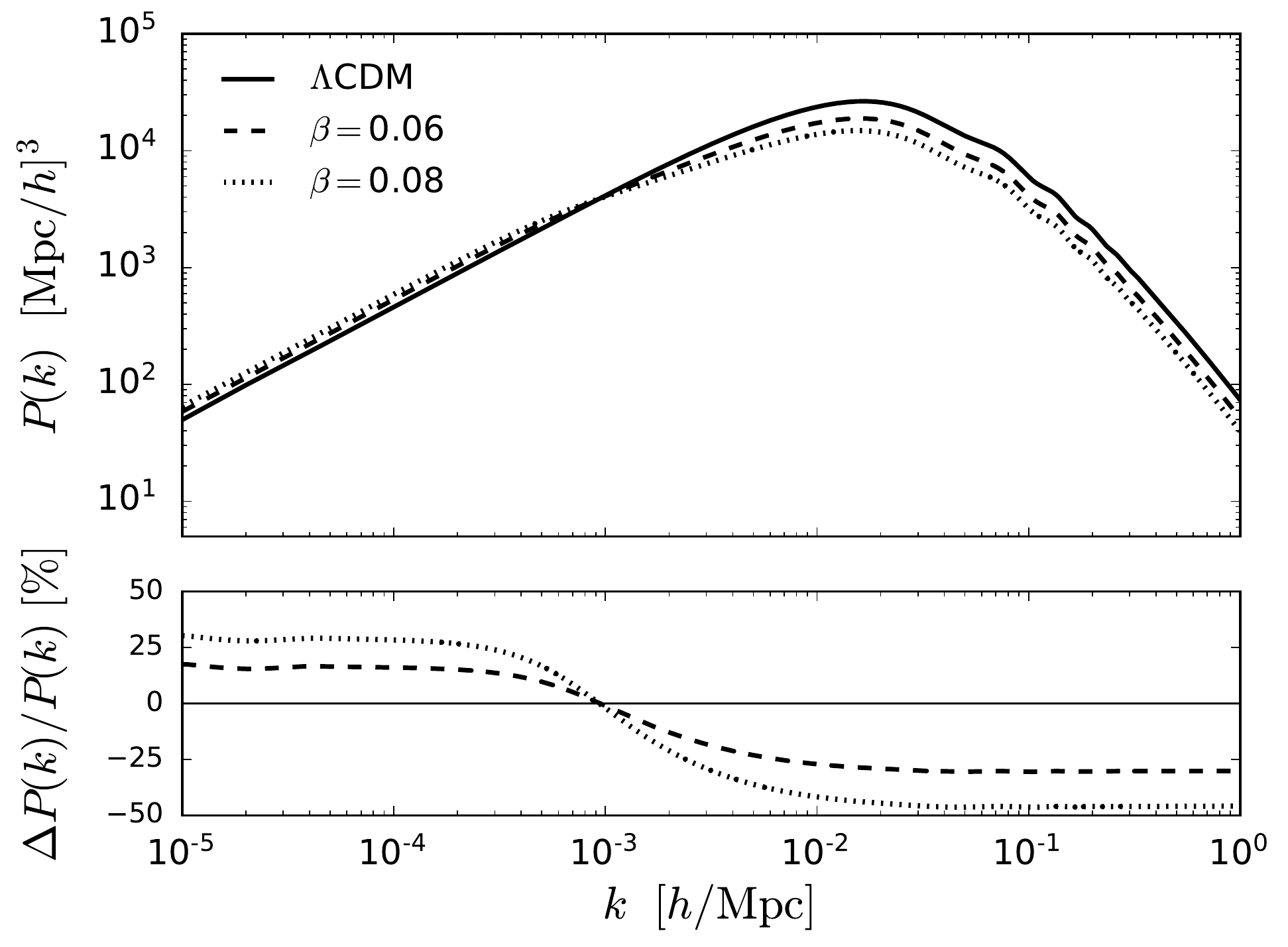}
    \caption{Linear matter power spectrum (at $z=0$) of CIDER for two values of $\beta$ and $\Lambda$CDM in absolute values (top panel) and relative differences to the $\Lambda$CDM one (bottom panel).}
    \label{fig:cq_linear_power_spectrum}
\end{figure}

\begin{figure}[b]
    \centering
    \includegraphics[scale=0.53]{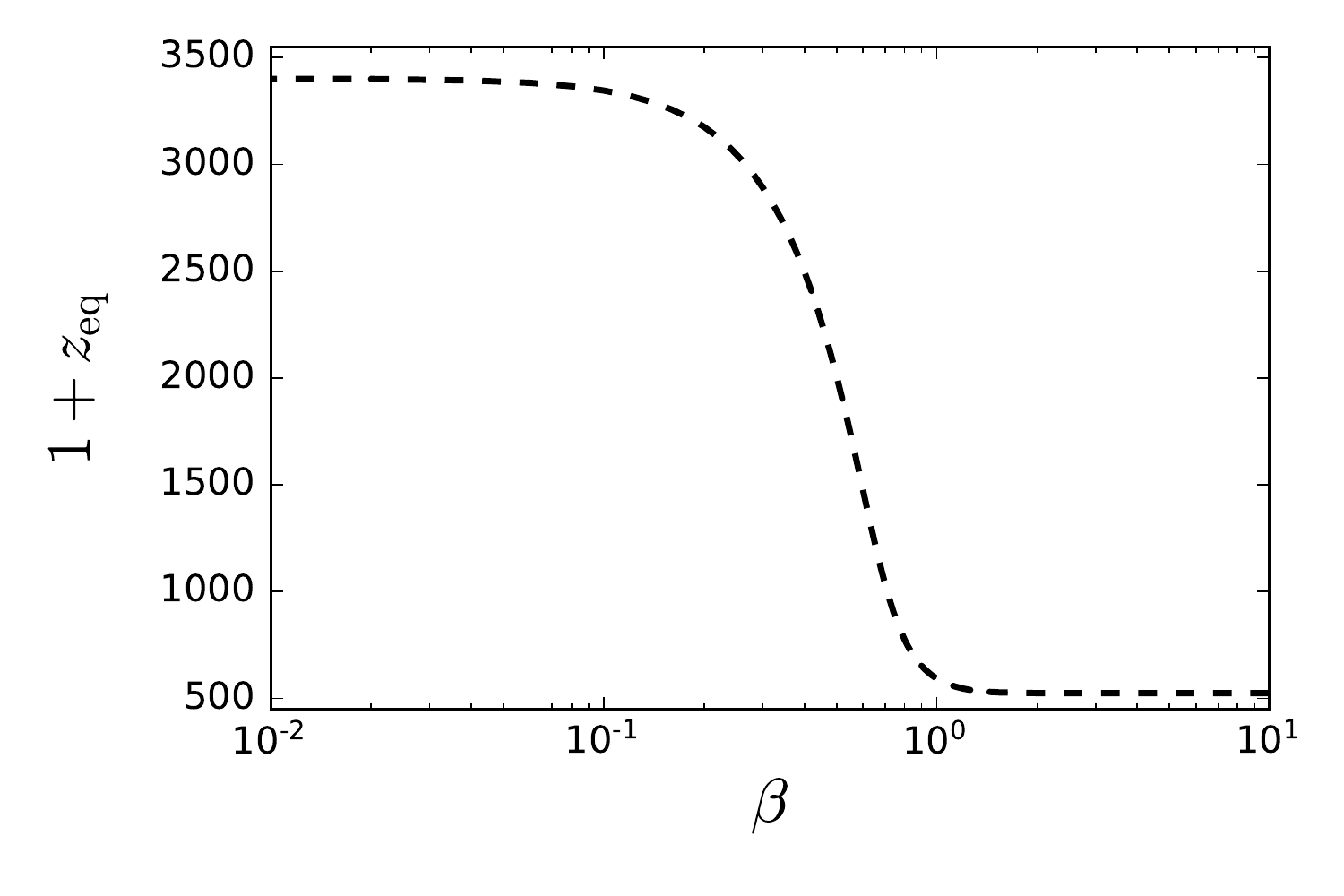}
    \caption{Variation of the matter-radiation equality redshift, $1+z_{\rm eq}$, as a function of the coupling $\beta$.}
    \label{fig:cq_z_eq}
\end{figure}

The linear matter power spectrum is depicted in the upper panel of Fig.~\ref{fig:cq_linear_power_spectrum}. 
We can identify an enhancement of the perturbations at the largest scales, which remain roughly constant, at $18\%$ and $30\%$ larger than their $\Lambda$CDM counterpart, for $\beta = 0.06$ and $\beta=0.08$ respectively. The importance of the analysis at such large scales might be argued since there are cosmic variance statistical limitations. Nonetheless, we may grasp the basis for such enhancement of the dark matter perturbations in this regime. We have shown that a stronger interaction leads to a smaller amount of dark matter (and larger amount of dark energy) throughout time. This will lead to delay the start of the matter dominated epoch. This trend is depicted in Fig.~\ref{fig:cq_z_eq}. The redshift of equality is given by,
\be\label{z_eq}
1+z_\texttt{eq}=\frac{\Omega_b^0+\Omega_{cdm}^0\,e^{-\kappa\beta\phi_{\rm eq}}}{\Omega_r^0}\mathcomma
\ee
which coincides with the $\Lambda$CDM one when $\beta=0$. Since $\beta\phi_\texttt{eq}\geqslant0$, the equality redshift decreases with $\beta$, as verified in Fig.~\ref{fig:cq_z_eq}. As $\beta$ grows the matter-radiation equality redshift asymptotically tends to the value $\Omega_b^0/\Omega_r^0$. In such case, dark matter is subdominant and baryons become the leading matter density. 

Since the matter-radiation equality is being delayed, the large super-Hubble scales, that are not caught by the expanding Hubble sphere during radiation domination, grow for a longer period of time with respect to the uncoupled case. Overdensities outside the Hubble sphere are not affected by radiation pressure thus are allowed to grow. A stronger interaction inevitably causes these super-Hubble modes to grow throughout a larger period of time, in comparison to $\Lambda$CDM. This is the dominant physical process leading to the relative enhancement of the matter power spectrum at the very large scales. Given that the interaction modifies the relative amplitude between scales, it changes the global tilt which depends on the primordial spectrum tilt, $n_s$. We can therefore expect to see $n_s$ decreasing as $\beta$ increases. Moreover, the interaction shifts the scales for which growth is suppressed by the sub-Hubble regime as the coupling also modifies the scale $k_\texttt{eq}$ of the power spectrum maximum determined by the time of equality. A perturbation mode on scale $k$ of physical size (wavelength) $2\pi a/k$ crosses the Hubble radius $1/H$ when $k\sim aH$. The scale of the maximum is then given by,
\be\label{z_eq2}
k_\texttt{eq}=\sqrt{2}H_0\frac{\Omega_b^0+\Omega_{cdm}^0e^{-\kappa\beta\phi_\texttt{eq}}}{\sqrt{\Omega_r^0}}\mathcomma
\ee
and decreases as $\beta$ increases (the peak is shifted to the left in Fig.~\ref{fig:cq_linear_power_spectrum}). The smaller scales that gradually enter the growing Hubble radius during the radiation era remain frozen for a longer period than in the standard model. Perturbations with {$k\gg k_{\rm eq}$} freeze as they are caught by the Hubble sphere since radiation pressure does not allow the clustering of matter. Subsequently, during the matter era, growth on such scales is further suppressed against a $\Lambda$CDM scenario. Ref.\cite{Barros:2018efl} found that dark matter overdensities inside the Hubble radius are suppressed, as a result of the interplay between the effective gravitational constant, ${G_{\rm eff}=G(1+2\beta^2)}$, induced by the fifth force, and the change in the amount of matter throughout the cosmic history. This last effect will be explained in more detail ahead.

\begin{figure}
    \centering
    \includegraphics[scale=.4]{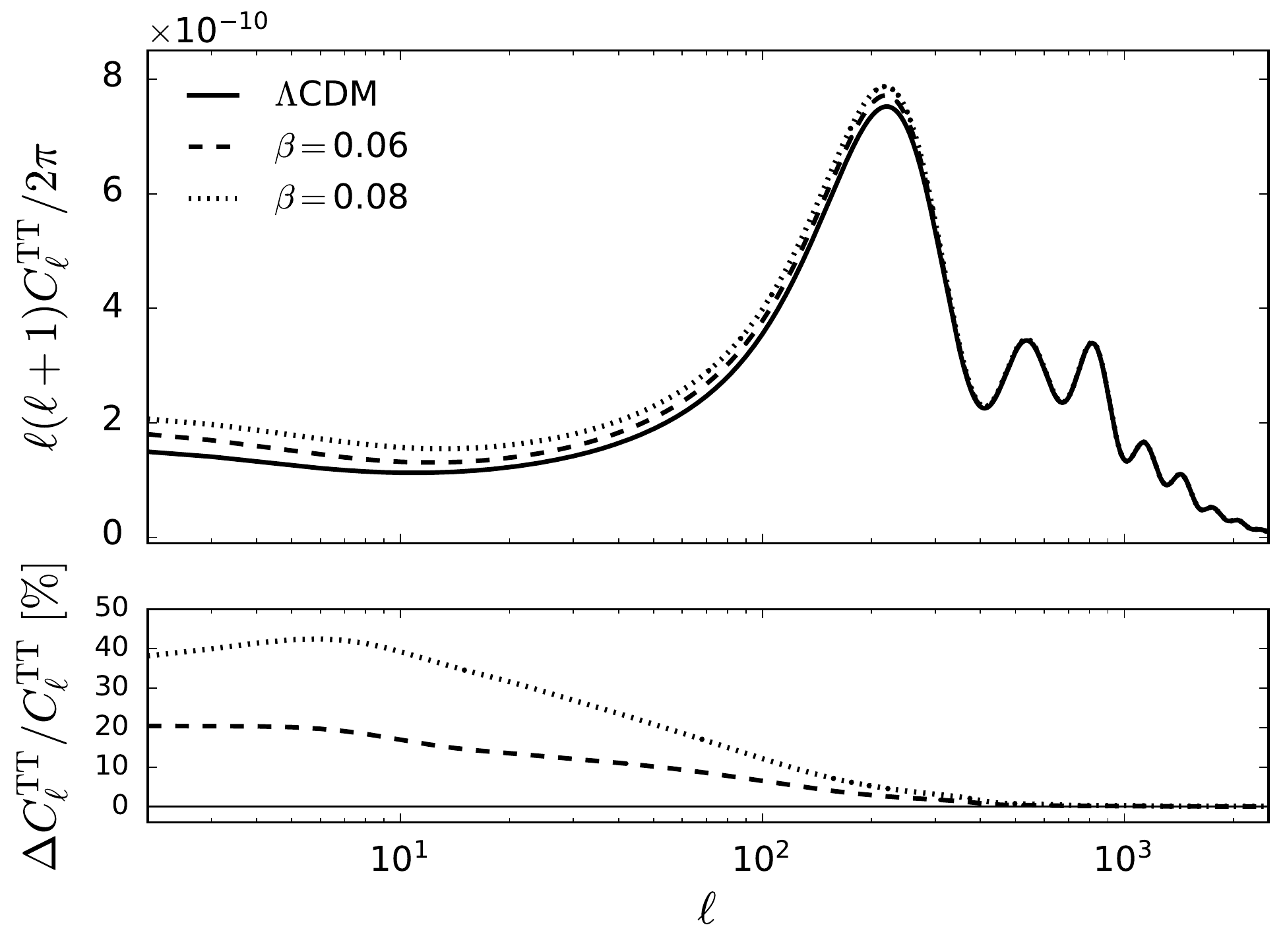}
     \caption{CMB power spectrum of CIDER for two values of $\beta$ and $\Lambda$CDM in absolute values (top panel) and relative differences to the $\Lambda$CDM one (bottom panel).}
    \label{fig:cq_cmb_tt}
\end{figure}

Regarding the CMB, the temperature angular power spectrum is shown in Fig.~\ref{fig:cq_cmb_tt}. The main deviations from the standard model case are at multipoles $\ell\lesssim 200$. In such a regime, the main physical processes influencing the radiation spectrum is the Integrated Sachs Wolf effect (ISW). We can discriminate between two main contributions due to this process. On the one hand, as the matter density decreases for larger couplings, the matter-radiation equality is shifted towards smaller redshifts. This will impact the redshift of photons at early times, when radiation was not negligible, thus contributing to the early ISW term, which manifests more prominently near the first acoustic peak that increases with $\beta$. On the other hand, at later times when the contribution of $\phi$-dark energy becomes non-negligible, the variation on the evolution of dark matter for different couplings directly affects the gravitational potentials via the Poisson equation. Dark matter only interacts with photons via gravity. Thus, photons travelling from the last scattering surface towards us will experience these gravitational potentials and any variation within them. The derivatives of the Bardeen potentials are directly related with the late time ISW effect term and dominate at small multipoles. The main effect leading to the enhancement of the CMB spectra low-$\ell$ tail, \textit{i.e.} ${\ell \lesssim 10}$, is this late time ISW effect. This is related to the variation of the gravitational potentials stemming from the variations on the total amount of matter throughout history with varying $\beta$. We notice that this enhancement reaches around $40\%$ at ${\ell\approx 10}$, for ${\beta = 0.08}$, to smaller multipoles.

\begin{figure}
    \centering
    \includegraphics[scale=0.4]{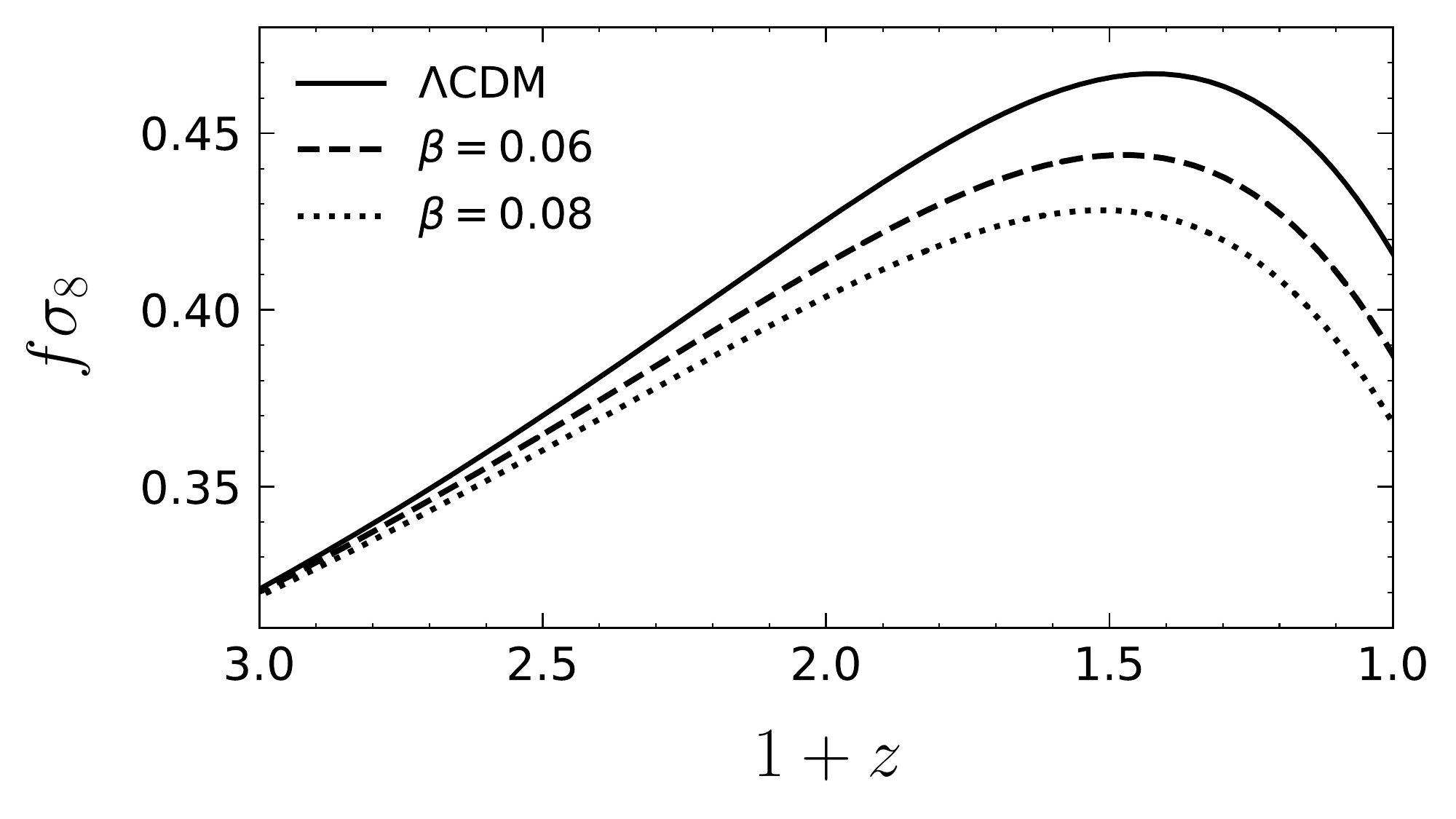}
    \caption{Values of the $f\sigma_8$ parameter as a funtion of redshift.}
    \label{fig:cq_fs8}
\end{figure}

The relative motion of galaxies within a cluster can leave imprints on cosmological data. Specifically, these peculiar velocities can squash the image of a cluster when plotted in redshift space. This effect is commonly known as redshift space distortions \cite{Kaiser:1987qv}. From such observations it is possible to grasp the physics of dark matter clustering through the growth rate parameter, ${f=d\ln \delta_m(a)/d\ln a}$, where we define a total matter density contrast, encompassing both CDM and baryons,
\be
\delta_m = \frac{\rho_c\delta_c+\rho_b\delta_b}{\rho_c+\rho_b}\mathperiod
\ee
Particularly, from RSD surveys, one is able to directly extract the value of the parameter
\be\label{fs8_eq}
f\sigma_8(a)=\frac{\sigma_8(0)}{\delta_m(0)}\frac{d\,\delta_m(a)}{d\ln a}\mathcomma
\ee
where ${\sigma_8(0)}$ is the present amplitude of the matter power spectrum at the scale of $8h^{-1}$Mpc. There might be different non-linear pattern signatures due to the impact of the coupling on the velocity fields. The non-linear regime would require a different treatment, such as N-body simulations (see \cite{Baldi:2022uwb}), which goes beyond the scope of the present work. The behaviour of $f\sigma_8$ for the CIDER model is depicted in Fig.~\ref{fig:cq_fs8} for different values of the coupling. Increasing the values of $\beta$ results in smaller values for $f\sigma_8$. This effect can be better appreciated by inspecting the equation governing the evolution of dark matter overdensities in the Newtonian limit \cite{Barros:2018efl},
\begin{gather}
\ddot{\delta}_c +\dot{\delta}_c\left( aH - \kappa\beta\dot{\phi} \right)-\frac{a^2\kappa^2}{2}\rho_c\delta_c\left(1+2\beta^2\right)\nonumber \\
-\frac{a^2\kappa^2}{2}\rho_b\delta_b=0\mathperiod \label{newt_limit}
\end{gather}
It is possible to identify two main differences from standard $\Lambda$CDM. An extra friction term, proportional to $\beta\dot{\phi}$, which is always negative (since ${\beta\dot{\phi}>0}$) and the emergence of a fifth force, induced by the scalar quintessence, with ${G_{\rm eff}/G = 1+2\beta^2}$. Both these terms source the growth of perturbations in the linear regime. However, the perturbations are suppressed for increasing values of $\beta$. This is because both effects described are subdominant when compared with the change on the background matter density, $\rho_c$, which decreases for stronger couplings (see lower panel of Fig.~\ref{fig:density_dilution}). This balances both the dragging and the fifth force terms, and results on an overall suppression of the matter clustering, thus slowing the growth of $f\sigma_8$ for increasing $\beta$ as seen in Fig.~\ref{fig:cq_fs8}.

%%%%%%%%%%%%%%%%%%%%%%%%%%%%%%%%%%%%%%%%%%%%%%%%%%%%%%%%%%%%%%%%%%%%%%
\section{Cosmological observations}\label{sec:obs}
%%%%%%%%%%%%%%%%%%%%%%%%%%%%%%%%%%%%%%%%%%%%%%%%%%%%%%%%%%%%%%%%%%%%%%

\subsection{Data}
We test the model with measurements of the angular power spectra of the CMB temperature and polarization in the early Universe, as well as with probes of structure formation in the late Universe, namely weak lensing and redshift space distortion. For the CMB we choose the Planck 2018 data \cite{Planck:2019nip} for which the likelihood codes are provided by the Planck Team\footnote{Likelihood downloaded from http://pla.esac.esa.int/pla}. The power spectrum is derived from Planck data combined with the nine-year WMAP sky maps \cite{2013ApJS..208...20B} on the large scales, and the 408-MHz survey \cite{1982A&AS...47....1H}, covering $93 \%$ of the sky. We carry out the statistical analysis with the Planck low-$\ell$ likelihood for the CMB measurement on the large scales, and the Planck high-$\ell$ lite likelihood for the smaller scales. For high-$\ell$, the lite likelihood version contains only one nuisance parameter, allowing us to have a faster convergence since we are working in a lower dimensional space.

As for the structure formation data, we select two samples: the weak lensing KiDS-450 dataset and the RSD Gold-2018 dataset. As far as KiDS is concerned, we use the shear power spectra and likelihood code provided in \cite{Kohlinger:2017sxk}. To compute the likelihood of our coupled quintessence model given the KiDS data, we vary the cosmological parameters values and also two main nuisance parameters that account for the bias in the measurements: the $\textit{A}_{\rm IA}$ parameter for the uncertainty in the amplitude of the intrinsic alignment effect and the $\textit{c}_{\rm min}$ parameter for the uncertainty on the dark matter power spectrum amplitude due to feedback from baryons. Also, the likelihood code already accounts by default for the uncertainty in the $n(z)$ distribution of galaxies by randomly choosing one of one thousand realisations. In the KiDS dataset the data points mostly lie in the non-linear regime of the matter power spectrum ($k \geqslant 0.2\,h{\rm{Mpc}}^{-1}$). For that reason we apply to the linear matter power spectrum the HMcode non-linear correction \cite{Mead:2020vgs} already implemented in CLASS and valid for coupled models that behave similarly to $\Lambda$CDM. Although this HMcode version was not specifically tailored to the CIDER model, it is expected to give optimized results on the small scales. It was especially tested with a similar conformal coupling between dark matter and quintessence, showing a few per cent accuracy of the power spectrum for $k<1\,h\rm{Mpc}^{-1}$ when compared to $N$-body simulations \cite{Mead:2016zqy}. In particular, the exponential potential tested is akin to the CIDER scalar field one, $V\propto\exp(-\phi/3\beta)$, which can be reconstructed during the deep matter dominated era to a good approximation \cite{Baldi:2022uwb}.

Finally, we use RSD in the clustering measurements of galaxy redshift surveys as a probe of structure growth, since galaxy peculiar velocities are produced by matter overdensities. Causing distortions along the line-of-sight, the peculiar velocities are responsible for the anisotropies in the observed correlation function. As they are measured in various redshift bins, one can obtain the evolution of the growth rate of matter perturbations, $f\sigma_8$. We use the publicly available likelihood\footnote{Likelihood downloaded from \\ https://github.com/snesseris/RSD-growth} which is based on the Gold-2018 compilation of 22 measurements of $f\sigma_8$ in Ref.~\cite{PhysRevD.102.103526}. The points in this dataset are unique and statistically robust \cite{PhysRevD.98.083543}.

\begin{table*}[t]
\begin{minipage}{\textwidth}
\begin{tabular} {l l l l l l l}
\hline
 & \multicolumn{2}{c}{\hspace{-0.9cm}Planck}  & \multicolumn{2}{c}{\hspace{-0.5cm}KiDS} & \multicolumn{2}{c}{\hspace{-0.3cm}RSD} \\[0.1cm]
  Parameter\hspace{0.5cm} & \multicolumn{1}{c}{\hspace{-0.5cm}CIDER}  & \multicolumn{1}{c}{\hspace{-1cm}$\Lambda$CDM}   & \multicolumn{1}{c}{\hspace{-1cm}CIDER}  & \multicolumn{1}{c}{$\Lambda$CDM}
  & \multicolumn{1}{c}{\hspace{-0.5cm}CIDER} & \multicolumn{1}{c}{$\Lambda$CDM}
  \\
\hline\\ [-0.3cm]

{\boldmath$|\beta|$} & $< 0.00995$ & n.a. & $0.087^{+0.032}_{-0.017}$ & n.a. & $0.079^{+0.058}_{-0.036}$ & n.a. \\[0.09cm]

{\boldmath$\Omega_{b }$} & $0.04889\pm 0.00069$ & $0.04876\pm 0.00069$ & $0.0425^{+0.0056}_{-0.0076}$ & $0.0401^{+0.0047}_{-0.0081}$ & --- & --- \\[0.09cm]

{\boldmath$\Omega_{c}$} & $0.2569\pm 0.0075$ & $0.2582\pm 0.0076$ & $0.250^{+0.056}_{-0.130}$ & $0.207^{+0.085}_{-0.11}$ & $0.175^{+0.037}_{-0.061}$ & $0.224^{+0.039}_{-0.049}$ \\[0.09cm]

{\boldmath$H_{0}$} & $67.76\pm 0.62$ & $67.88\pm 0.62$ & $73.0\pm 4.8$& $> 72.8$ & --- & ---\\ [0.09cm]

{\boldmath$n_{s }$} & $0.9699\pm 0.0043$ & $0.9702\pm 0.0043$ & $0.906^{+0.059}_{-0.200}$& $1.02\pm 0.13$ & --- & --- \\[0.09cm]

{\boldmath$\ln 10^{10}A_{s }$} & $3.1218\pm 0.0059$ & $3.1217\pm 0.0058$ & $>3.17$ & $< 3.82$ & $3.74\pm 0.70$ & $2.96^{+0.44}_{-0.50}$ \\[0.09cm]

{\boldmath$\Omega_{m}$} & $0.3088\pm 0.0083$ & $0.3084\pm 0.0083$ & $0.294^{+0.056}_{-0.130}$ & $0.249^{+0.084}_{-0.12}$ & $0.218^{+0.039}_{-0.060}$ & $0.267^{+0.038}_{-0.048}$\\[0.09cm]

{\boldmath$\sigma_8$} & $0.8370^{+0.0067}_{-0.0056}$ & $0.8399\pm 0.0053$ & $0.84\pm 0.18$ &  $0.85^{+0.17}_{-0.22}$ & $0.873^{+0.073}_{-0.095}$ &$0.795^{+0.042}_{-0.050}$ \\[0.09cm]

{\boldmath$S_8$} & $0.849\pm 0.017$ & $0.852\pm 0.016$ & $0.796\pm 0.064$ & $0.737^{+0.038}_{-0.031}$ & $0.733\pm 0.038$ & $0.744\pm 0.040$\\[0.1cm]

\hline
{\boldmath$\Delta\chi^2_{\rm{red}}$} &  \multicolumn{2}{c}
{\hspace{-0.9cm}$0.0025$}  & \multicolumn{2}{c}{\hspace{-0.5cm}$0.0034$} & \multicolumn{2}{c}{\hspace{-0.3cm}$0.0449$} \\[0.09cm]
{\boldmath$\ln B_{\phi\Lambda}$}& \multicolumn{2}{c}{\hspace{-0.9cm}$-2.915$}  & \multicolumn{2}{c}{\hspace{-0.5cm}$0.329$} & \multicolumn{2}{c}{\hspace{-0.3cm}$-0.141$} \\[0.09cm]
\hline
\end{tabular}
\caption{Mean and 68\% uncertainty estimates of the 6 basis parameters and 3 derived parameters for the CIDER and $\Lambda$CDM models by the 3 data sets. The parameters not constrained are indicated with '---'. The statistics used for model comparison, ${\Delta\chi^2_{\rm{red}}\equiv\chi^2_{\rm{red}}(\rm{CIDER})-\chi^2_{\rm{red}}(\Lambda\rm{CDM})}$ and $B_{\phi\Lambda}\equiv B(\rm{CIDER})/B(\Lambda\rm{CDM})$, are also shown.}
\label{table}
\end{minipage}
\end{table*}

\subsection{Setup}
To infer the parameters, we use the Nested Sampling algorithm \cite{Skilling2006} of the Multinest library \citep{Feroz_2009} wrapped with PyMultiNest \cite{Buchner:2014nha} in the MontePython package \citep{MP1}. This package is prepared to work integrated with our modified version of CLASS and already contains several likelihood codes for the most recent experiments. The resulting Monte Carlo samples are analysed with the GetDist package \cite{Lewis:2019xzd}.

We test the CIDER model with six free parameters, one parameter for the coupling ($\beta$) and five other fundamental cosmological parameters: the primordial power spectrum amplitude ($A_s$) and slope ($n_s$), the baryon and dark matter densities (${\omega_b = \Omega_b h^2}$ and ${\omega_c = \Omega_c h^2}$), and the reduced Hubble constant ($h$). We also perform the same statistical analysis with the $\Lambda$CDM model for the sake of model comparison. Note that this corresponds to the standard six parameter cosmological model studied by Planck for $\Lambda$CDM, minus the reionization parameter $\tau_{\textrm{reio}}$ not relevant here.

\subsection{Likelihood analysis}

The constraints obtained with the three likelihood analyses are shown in Table~\ref{table} for the basis and derived parameters $\Omega_m$, $\sigma_8$ and ${S_8=\sigma_8\sqrt{\Omega_m/0.3}}$. Results from the same three likelihood analyses performed for the $\Lambda$CDM model are also given. We show in Appendix \ref{appendix} the 2D projections and probability distributions of the CIDER parameters.

The expected decrease of the spectral index discussed in the previous section is well visible in the KiDS data, reducing from ${n_s=1.02\pm 0.13}$ ($\Lambda$CDM model) to ${n_s= 0.906^{+0.059}_{-0.200}}$ (CIDER).

\begin{figure*}
    \centering
    \includegraphics[scale=0.48]{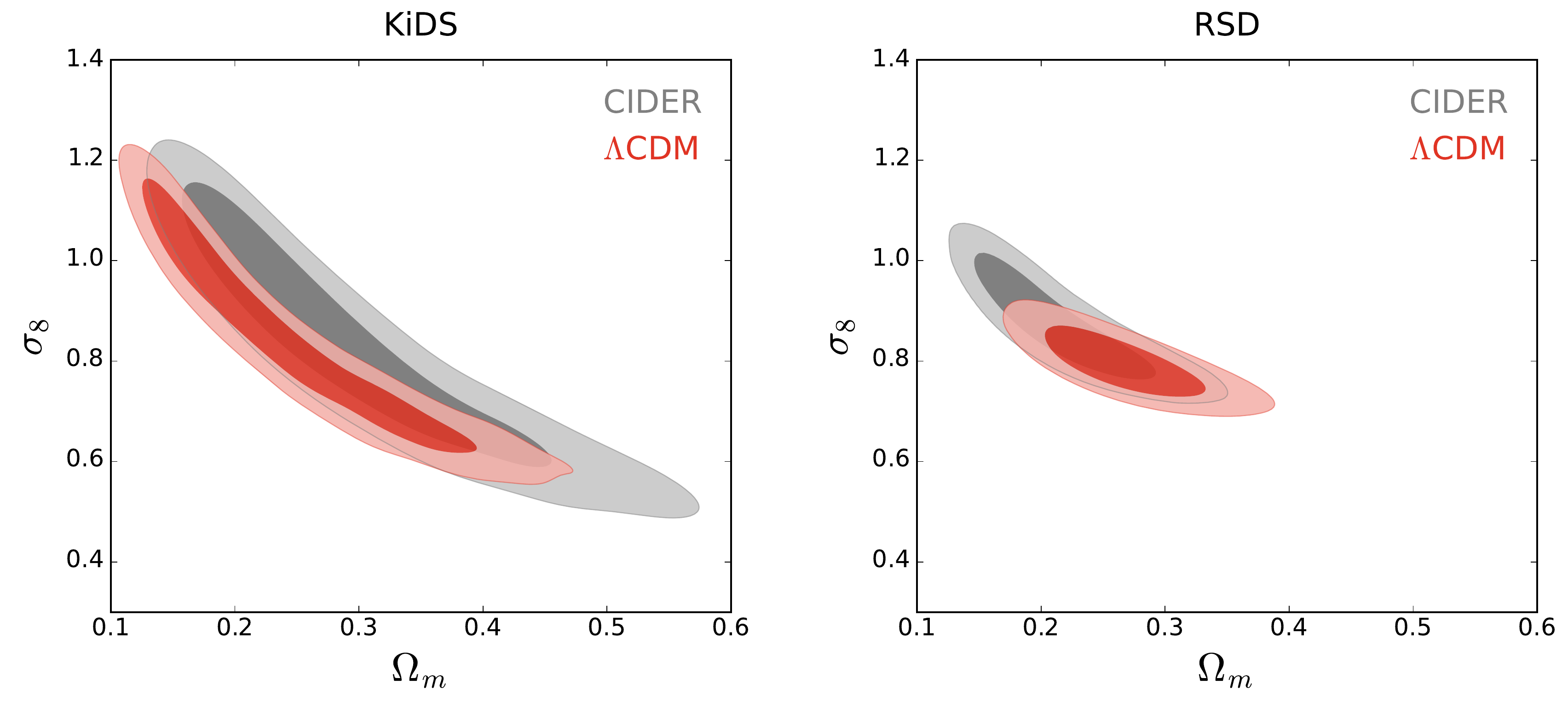}
    \caption{Comparison between $\Lambda$CDM and CIDER constraints. Marginalised contours (68\% and 95\% confidence levels) from KiDS (left panel) and RSD (right panel) analyses.}
    \label{fig:lcdm_cq_bis}
\end{figure*}

\begin{figure*}
    \centering
    \includegraphics[scale=0.48]{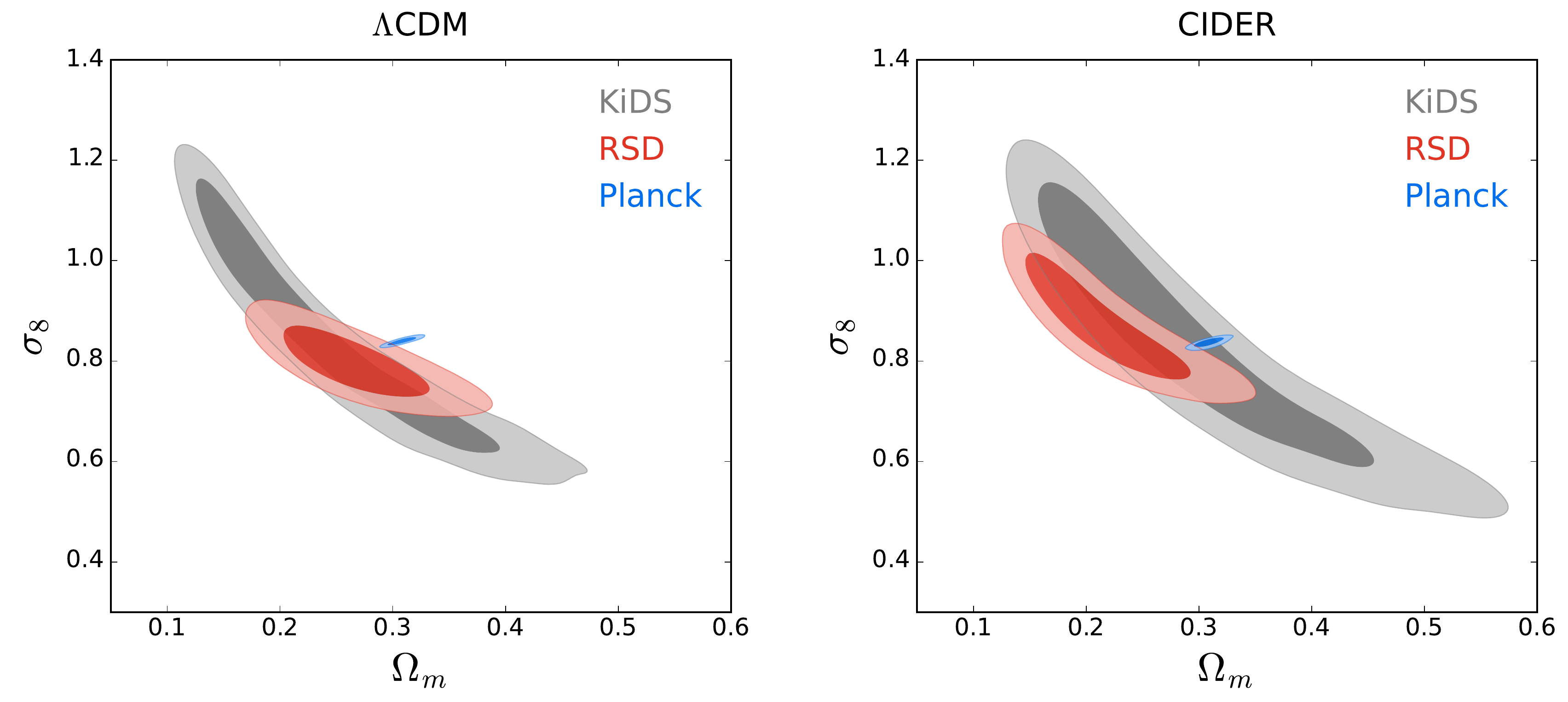}
    \caption{$\sigma_8$ tensions between the different data sets. Marginalised contours (68\% and 95\% confidence levels) from the three analyses for $\Lambda$CDM (left panel) and CIDER (right panel).}
    \label{fig:lcdm_cq}
\end{figure*}

The presence of the additional parameter $\beta$ enlarges the confidence regions in the parameter space when compared to $\Lambda$CDM, particularly in the ${(\Omega_m,\sigma_8)}$ plane where the contours are widened to the right as shown on the left panel of Fig.~\ref{fig:lcdm_cq_bis}. The most predominant effect of $\beta$ on the matter power spectrum is the growth suppression of the small scales. To compensate for the existence of the interaction, one can expect an increase in $S_8$ together with its uncertainty. This is confirmed in Table~\ref{table}, where the KiDS estimated $\sigma_8$ decreases slightly and $\Omega_m$ increases significantly, resulting in an increase in $S_8$ from ${0.737^{+0.038}_{-0.031}}$ to ${0.796\pm 0.064}$. In this way, the coupled quintessence model seems to relax the $\sigma_8$ tension that exists between KiDS and Planck in the $\Lambda$CDM model, as illustrated in Fig.~\ref{fig:lcdm_cq}.

\begin{figure*}
    \centering
    \includegraphics[scale=0.8]{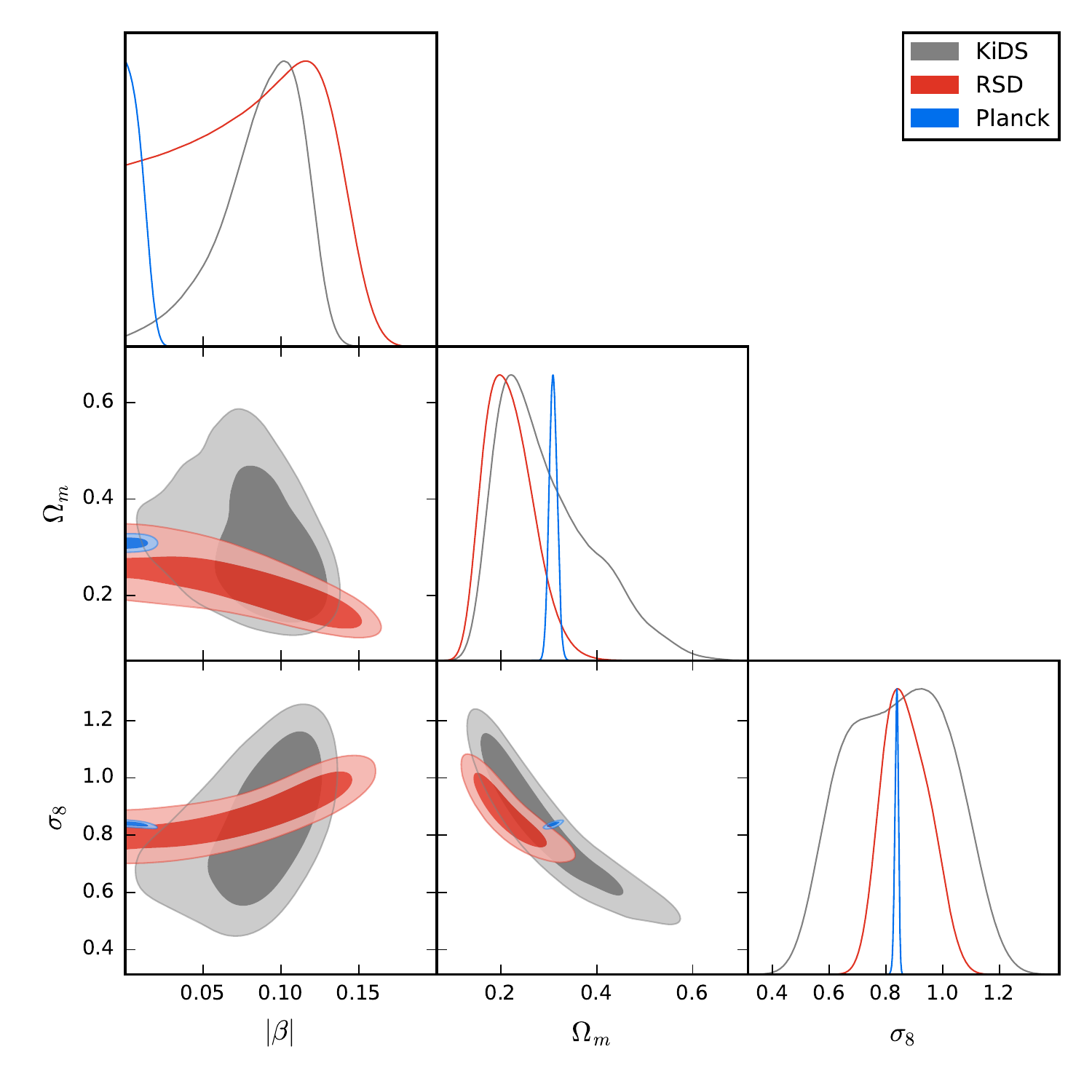}
    \caption{Marginalised probability distribution and marginalised contours (68\% and 95\% confidence levels) for the three analyses of the CIDER model.}
    \label{fig:contours}
\end{figure*}

However, this tension is somehow transferred to the parameter $\beta$ which vanishes according to Planck, in contradiction with the KiDS result that suggests ${|\beta|=0.087^{+0.032}_{-0.017}}$ (see Fig.~\ref{fig:contours}). The dominant effect of the interaction on the CMB power spectrum is the large-scale increase. There is little room to compensate for this effect by varying other parameters without leading to a change in the first peak. It appears therefore that the only good fit to the Planck observations is achieved for a vanishing coupling.

As for the RSD data, the tension with Planck is only marginally alleviated by the existence of the coupling as the confidence contours are enlarged towards higher values of $\sigma_8$ and lower values of $\Omega_m$ (see Figure~\ref{fig:lcdm_cq_bis}, right panel). Unlike the KiDS result, the interaction significantly increases $\sigma_8$ and decreases $\Omega_m$, resulting in a slight decrease in the RSD $S_8$ estimate (see Table~\ref{table}). We note that $\sigma_8$ and $\Omega_m$ have identical and scale-independent effects in weak lensing measurements since $\Omega_m$ mainly changes the amplitude of the lensing through the Poisson's equation. On the contrary, in RSD, which is a function of redshift and not scale, $\Omega_m$ has a greater effect at low redshift through the growth factor $f$ so that $\Omega_m$ and $\sigma_8$ do not compensate as in the weak lensing case.

\subsection{Model comparison}

The results of the $\Lambda$CDM likelihood analyses are shown in Table~\ref{table}. We perform a CIDER vs. $\Lambda$CDM model comparison based on the minimum $\chi^{2}_{\rm red}$ values and on the Bayes factors.

The minimum $\chi^{2}_{\rm red}$ is the best-fit $\chi^2$ per degrees of freedom, with the number of degrees of freedom being computed as the number of data points minus the number of model parameters used in the analysis. The positive $\Delta\chi^{2}_{\rm red}$ in Table~\ref{table} indicate a better fit for the $\Lambda$CDM model in every experiment. However, since the $\chi^{2}_{\rm red}$ values are similar in all cases, we do not see a significant disadvantage in the CIDER model despite its additional parameter, $\beta$.

The usage of the Nested sampling algorithm has the advantage of also providing the Bayesian evidence of every analysis, allowing us to compute the Bayes factor \cite{Trotta_2008}, $\ln B_{\phi\Lambda}$, and perform a model comparison based on the full likelihood rather than on the best-fit only. Applying the Jeffrey's scale (see \cite{Trotta_2008}) to the Bayes factors shown in Table~\ref{table}, we find that KiDS and RSD data are inconclusive. Planck data moderately prefer $\Lambda$CDM, in agreement with the $\chi^{2}_{\rm red}$ comparison. This result suggests that the coupling between dark matter and dark energy is not favored by high redshift data, or that if there is such a coupling, it should be significantly close to zero at the time of the last scattering. 

%%%%%%%%%%%%%%%%%%%%%%%%%%%%%%%%%%%%%%%%%%%%%%%%%%%%%%%%%%%%%%%%%%%%%%
\section{Conclusions}\label{sec:conclusions}

We have studied the constrained interacting dark energy (CIDER) model specifically tailored to satisfy the background constraints given by cosmological probes. This is done by taking note of the success of the concordance $\Lambda$CDM model and fixing the quintessence scalar potential to give us the same background evolution, \textit{i.e.} 
${H = H_{\Lambda{\rm CDM}}}$. This ad hoc assumption can be motivated by the fact that small deviations on the background cosmology may have a large impact at linear level (such as a shift on the position of the first CMB peak).

This work complements two previous articles \cite{Barros:2018efl,Barros:2019hsk} by analysing details on the background cosmology and capturing effects at the linear level that were still undisclosed. Through suitable approximations in the Klein-Gordon equation we were able to find an analytical expression for the scalar field evolution during radiation domination, where its energy density dilutes as ${\rho_{\phi}\propto a^{-2}}$. It was found that the radiation-equality epoch is shifted towards later times for increasing values of the coupling $\beta$. Due to this behaviour we have a prolonged growth of large mode matter overdensities leading to an enhancement of the matter power spectrum at large scales. Regarding the CMB temperature angular power spectrum, modifications on the dark gravitational potentials both at late times and near the photon decoupling era result on a modified ISW effect in contrast with the uncoupled ($\Lambda$CDM) model. This gives rise to an enhancement at large-$\ell$ on the CMB power spectrum. 

We then tested the CIDER model against late time weak lensing and redshift space distortion data, and early time measurements of the CMB temperature and polarization power spectra. The KiDS data seem to alleviate the current $\sigma_8$ tension between weak lensing and the Planck data since the $\beta$ parameter is able to broaden the confidence regions of the parameter space. The influence of the interaction on the physics of the matter power spectrum results on weak lensing favouring a lower value for the primordial tilt $n_s$. On the other hand, the RSD observable is not scale dependent so the dependence on the cosmological parameters is not identical to KiDS. We found that RSD data predicts a smaller value for $S_8$ in comparison with the base $\Lambda$CDM model, and the tension with Planck is only marginally alleviated.

The {\it a priori} assumption of fixing the background expansion rate, through Eq.~\eqref{assumption}, comes with subtle differences, when testing the model against observations, in contrast with standard coupled quintessence. The standard theory of conformally coupled scalar field dark energy \cite{Gomez-Valent:2020mqn,Pettorino:2012ts,Amendola:2011ie,Pettorino:2013oxa} presents one more free parameter, besides the coupling, related to the stiffness of the potential, usually denoted by $\lambda$, allowing a wider range of values for the coupling to be allowed by observations. In contrast, imposing this background evolution with only one extra free parameter results in stringent constraints over the dark interaction, suggesting negligible values for the coupling in this current model. Although large scale observations of the late Universe allow non-zero values for couplings between the dark species, the sensitivity of the Planck mission forbids large deviations in the physics of the early Universe, thus constraining the dark interaction to be negligible at such epoch. We indeed corroborate this behaviour with the Planck data favouring a vanishing coupling. This suggests that in order to avoid incompatibility with early Universe data, in particular by the Planck mission, one is compelled to seek models that hide the presence of the coupling throughout the early history of the Universe, with the interaction kicking in only at late times.
%%%%%%%%%%%%%%%%%%%%%%%%%%%%%%%%%%%%%%%%%%%%%%%%%%%%%%%%%%%%%%%%%%%%%%

\acknowledgments
The authors thank Marco Baldi for the careful reading and helpful comments, and the anonymous referee for the suggestions that improved the draft manuscript. B.J.B. is supported by the South African NRF Grants No. 120390, reference: BSFP190416431035; No. 120396, reference: CSRP190405427545. D.C. acknowledges support from IDPASC through the grant No. PD/BD/150489/2019. V.d.F. acknowledges FCT support under the grant reference 2022.14431.BD. This work was financed by FEDER -- Fundo Europeu de Desenvolvimento Regional -- funds through the COMPETE 2020 -- Operational Programme for Competitiveness and Internationalisation (POCI) -- , and by Portuguese funds through FCT -- Funda\c c\~ao para a Ci\^encia e a Tecnologia -- under projects PTDC/FIS-AST/28987/2017, PTDC/FIS-AST/0054/2021 and EXPL/FIS-AST/1368/2021, as well as UIDB/04434/2020 \& UIDP/04434/2020, CERN/FIS-PAR/0037/2019, PTDC/FIS-OUT/29048/2017.

\newpage

\appendix
\section{}\label{appendix}
\begin{figure}[h!]
    \centering
    \includegraphics[scale=0.30]{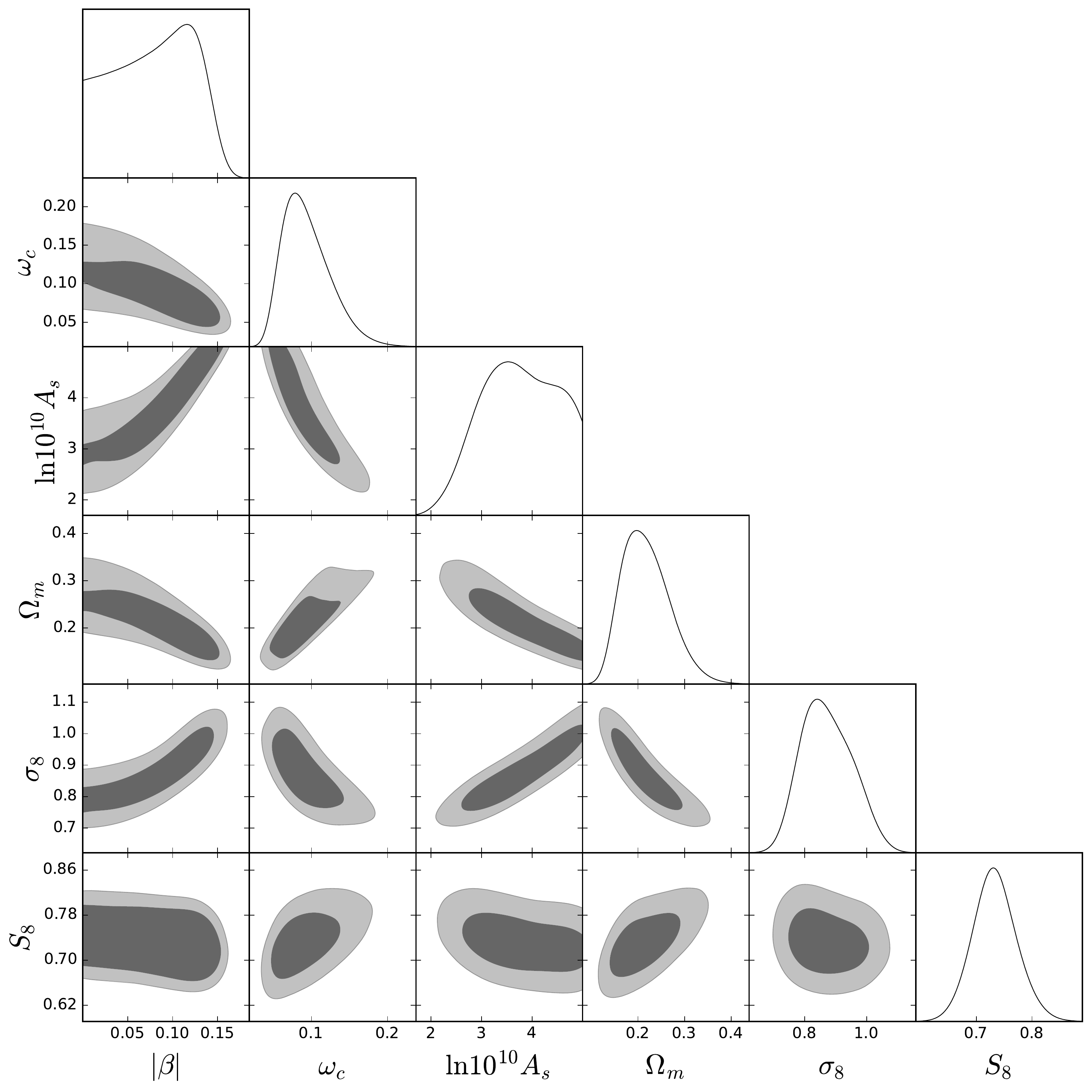}
    \caption{RSD marginalised probability distribution and marginalised contours (68\% and 95\% confidence levels) of the CIDER model.}
    \label{fig:ap_rsd}
\end{figure}

\begin{figure*}
    \centering
    \includegraphics[scale=0.32]{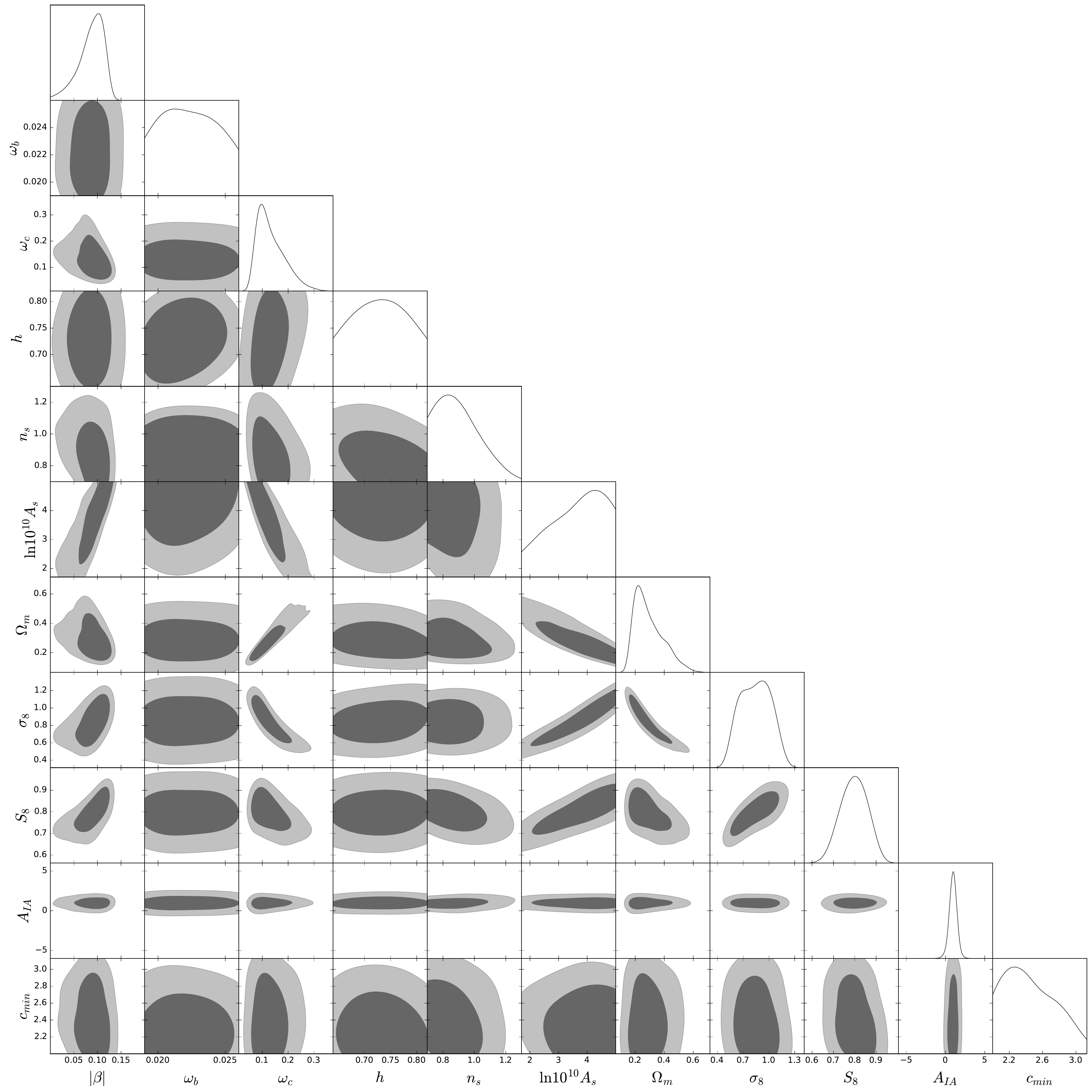}
    \caption{KiDS marginalised probability distribution and marginalised contours (68\% and 95\% confidence levels) of the CIDER model.}
    \label{fig:ap_kids}
\end{figure*}

\begin{figure*}
    \centering
    \includegraphics[scale=0.35]{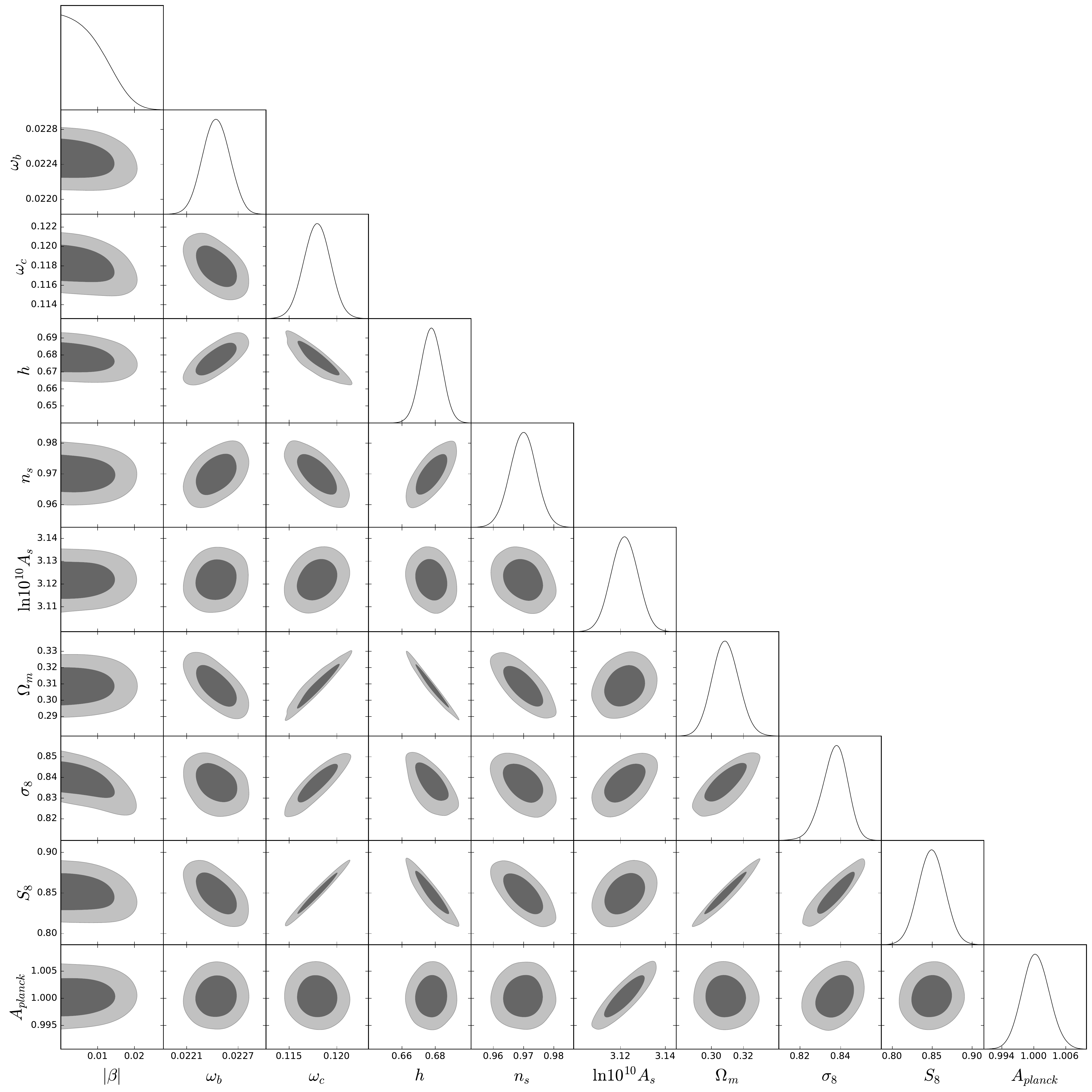}
    \caption{Planck marginalised probability distribution and marginalised contours (68\% and 95\% confidence levels) of the CIDER model.}
    \label{fig:ap_planck}
\end{figure*}

In this appendix we show the marginalised probability distributions and contours of the CIDER parameters for the three experiments used in the Bayesian inference: RSD in Fig.~\ref{fig:ap_rsd}, KiDS in Fig.~\ref{fig:ap_kids}, and Planck in Fig.~\ref{fig:ap_planck}.

\clearpage

\bibliography{bib}

\end{document}